# MODELING THE ADAPTION RULE IN CONTEXT-AWARE SYSTEMS

Mao Zheng[1], Qian Xu[2] and Hao Fan[3]

[1]Department of Computer Science, University of Wisconsin-LaCrosse, LaCrosse, USA
mzheng@uwlax.edu
[2]Amazon, Seattle, USA
xqia@amazon.com
[3]School of Information Management, Wuhan University, Wuhan, China
hfan@whu.edu.cn

*ABSTRACT*

*Context awareness is increasingly gaining applicability in interactive ubiquitous mobile computing systems. Each context-aware application has its own set of behaviors to react to context modifications. This paper is concerned with the context modeling and the development methodology for context-aware systems. We proposed a rule-based approach and use the adaption tree to model the adaption rule of context-aware systems. We illustrate this idea in an arithmetic game application.*

## 1. INTRODUCTION

Our world gets more connected everyday. These connections are driven in part by the changing market of smartphones and tablets. Pervasive computing environments are fast becoming a reality. The term "pervasive", introduced first by Weiser [1], refers to the seamless integration of devices into the user's everyday life. One field in the wide range of pervasive computing is the so-called context-aware system. Context-aware systems are able to adapt their operations to the current context without an explicit user intervention and thus aim at increasing usability and effectiveness by taking environmental context into account. Each context-aware application has its own set of behaviors to react to context modifications. Hence, every software engineer needs to clearly understand the goal of the development and to categorize the context in the application. We incorporate context-based modifications into the appearance or the behavior of the interface, either at the design time or at the run time. In this paper, we present application behavior adaption to the context modification via a context-based user interface in a mobile application, arithmetic game. The application's mobile user interface (MUI) will be automatically adapted based on the context information.

The user interface (UI) can include many features such as font color, sound level, data entry, etc. Every feature has some variables. For example, data entry can be done using typing, voice and tapping. From the designer's perspective, the adaptability of these features is planned either at the design time or during the runtime. Through the literature study, we proposed a rule-based approach model, and used an adaption tree to present this model. The adaption tree is what we named in our methodology. It is based on the extension of a decision table, the decision tree. We use the adaption tree to represent the adaption of the mobile device user interface to various context information. The context includes the user's domain information and dynamic environmental changes. Each path in the adaption tree, from the root to the leaf, presents an adaption rule. To illustrate our methodology, we implemented a context-aware application in the Android platform, the arithmetic game application.

There are two major platforms in the mobile device community: iOS and Android. This project chose Android development mainly for the reason of its openness. In addition, all the tools in the Android development are free and no special hardware is required.

The rest of the paper is organized as follows: in Section 2 we compare how our views are similar to other researchers and how they are different. Section 3, we briefly describe the arithmetic game application. Section 4 presents the rule-based approach and the fundamental concepts of the adaption tree. Section 5 discusses the development of the arithmetic game based on the adaption tree. Section 6 concludes the paper and outlines the directions of our on-going research.

## 2. Related Work

Some researchers define context as the user's physical, social, emotional or informational state, or as the subset of physical and conceptual states of interest to a particular entity [2]. The authors in [2] have presented the definition or interpretation of the term by various researchers, including Schilit and Theimer [3], Brown *et al.* [4], Ryan *et al.* [5], Dey [6], Franklin & Flaschbart [7], Ward *et al.* [8], Rodden *et al.* [9], Hull *et al.* [10], and Pascoe [11]. In Dey and Abowd [2], the authors are interested in context-aware systems, and so they focused on characterizing the term itself. In Pascoe [11], the author's interest is wearable computers, so his view of context is based on environmental parameters as perceived by the senses. Our work depends on the internal sensors of a mobile device, and the adaption of the mobile user interface features for both entering and accessing data. Our model is based on separating how context is acquired from how it is used, by adapting the mobile user interface features to the user's context.

Most of the research in this area has been based on analyzing context-aware computing that uses sensing and situational information to automate services, such as location, time, identity and action. More detailed adaption has been generally ignored. For example, input data based on context. In our research, we attempted to build the user's characteristics from both domain experience and mobile technology experience, and to collect all the context values corresponding to the user's task and then to automatically adapt the mobile user interfaces to the context information.

The process of developing context-based user interface has been explored in a number of other projects. Clercks *et al.* [12], for example, discuss various tools to support the model-based approach. Many studies have been conducted on adaption using a decision table. In [13], an approach is proposed for modeling adaptive 3D navigation in a virtual environment. In order to adapt to different types of users, they designed a system of four templates corresponding to four different types of users. Our work differs in that our adaption technique is based on composite context information that extracts values from sensors in smartphones and relates with the user's domain and mobile technology experiences. Then we develop a set of rules for the mobile user interface adaption. We used the adaption tree to model the context information and represent adaption rules. It also serves as the model for the design and implementation.

## 3. The Arithmetic Game Application

The arithmetic application is developed for users of different ages, with different arithmetic skills. The mobile user interface will adapt to the user's profile, actual performance and current time and local weather information. The device's orientation will also be discussed as one of the context information.

Users are required to register and obtain an account to login. During the user's registration process, the user's age is stored as one of the logic context information that will be used in the beginning of the app to assign the appropriate question level to the user.

There are two modes in the application: standard mode and review mode. The standard mode is to let the user practice or test their arithmetic skills through questions in different levels and units. The review mode is to let the user redo the questions he/she made mistakes on before.

In the standard mode, there are three levels corresponding to three age domains. Each level is divided into 10 units and each unit contains 10 problems. When a user answered all questions correctly in the last unit, or answered more than or equal to 90 questions correctly in the last level, the user can choose to level up, otherwise the user will stay at the same level.

When the user logs in for the first time, he/she is automatically assigned to a level based on the age information. All the problems are generated randomly based on the rules shown in the Table 1 below. Generating questions randomly instead of retrieving questions from a database, can avoid the users remembering the order of answers when they play the game again and again.

Table 1: Question Characteristics of Different Levels

| Age | Level | Operands for + | Operands for - | Operands for * | Operands for / |
|---|---|---|---|---|---|
| [0, 5] | 1 | Both [0, 10] | Both [0, 10] minuend > subtrah | Not available | Not available |
| [6, 12] | 2 | Both [0, 50] | Both [0, 50] minuend > subtrah | Both[0, 10] | Both [0, 10] Quotient is integer |
| >= 13 | 3 | Both [-100, 100] | Both [-100, 100] | Both[-10, 10] | Both [-100, 100] Quotient is integer |

When the users answer questions, there are 10 seconds for each question. A graphic countdown timer should work as a reminder.

The accuracy of the last unit is divided into three groups: [0%, 60%], (60%, 90%), [90%, 100%]. The arithmetic game presents three themes for three different accuracy groups respectively. For the first group, whose last unit accuracy is between 0% and 60% inclusively, the application simply presents a default theme. For the second group, whose last unit accuracy is between 60% and 90% exclusively, the user is able to design their own theme with their preferred color. In the application's setting, the user can choose preferred colors for different UI widgets. For the third group, whose last unit accuracy is between 90% and 100% inclusively, the user can design their own theme with their preferred color, local time and local weather. The weather icon follows the local weather information. The background image follows the local time period by default. During 6:00 – 17:00, the picture of a daytime scene is displayed as the background image; during 17:01 – 19:00, the picture of a sunset is displayed as the background image; during 19:01 – 5:59, the picture of a nighttime scene is displayed as the background image. The user can also choose to close the "time-based background image" option in the setting, thus the background will be presented in color style.

## 4. Rule-based Approach

Our work depends on the internal sensors of a mobile device, the user profile and the user's task. The key point of the approach is to capture and represent the knowledge required for the mobile user interface to automatically adapt to dynamics at run time, or to implement the adaptions at design time. The rule-based approach representation is what we are proposing. Figure 1 below shows our proposed approach.

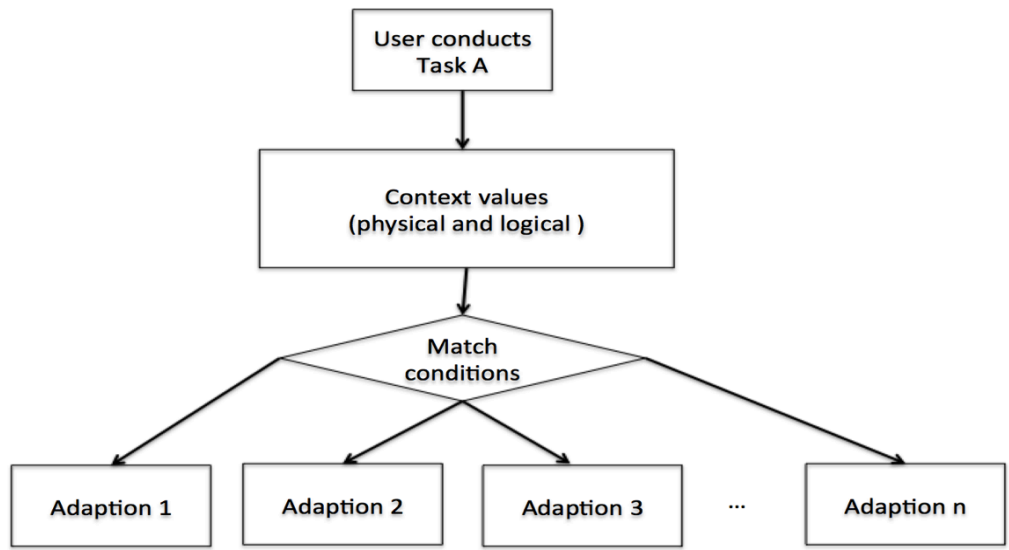

Figure1. Rule-based Approach

We use the adaption tree in this paper to show our rule-based approach in the context-based UI adaptation. The inspiration comes from mHealth [14] where the decision table was selected to represent the adaption rules of Mobile User Interfaces. However, in the process of adapting the decision table to our applications, we met some challenges; 1) the sequence is not clearly shown in the decision table, 2) the decision table method is useful for those applications that include several independent relationships among the input parameters, but it does not consider the relationships among the conditions, such as overlapping or redundancy, 3) a decision table does not scale up very well – when there are *n* rules, there are $2^n$ rules. Since each condition needs to be evaluated as true, false, or not applicable. It is not easy to make a full description for those conditions when *n* increases. Because of those limitations within the decision table, we presented our UI adaption rule using an adaption tree instead of a decision table.

The concept of adaption tree comes from the decision tree. It is a graphical representation of a decision-making situation. Compared to a tabular decision table, it takes up more room, but it shows the order of evaluating the conditions.

A decision tree is a flowchart-like structure in which each internal node represents a "test" on an attribute (e.g. whether a coin flip comes up heads or tails), each branch represents the outcome of the test and each leaf node represents a class label (decision taken after computing all attributes). The paths from root to leaf represent classification rules [15].

### 3.1 Adaption Tree Used in Our Research

We call the decision tree used in our project an "adaption tree," and our approach of adaption is to change the UI based on context. Below are some basic concepts used in our research.

**UI feature:** UI features are the smallest atomic unit for describing UI content on a mobile device. Table 2 shows some UI features and their actions (also known as values) below:

Table 2: UI features and their actions

| UI features | Action(values) |
| --- | --- |
| Font size | Small, medium, large |
| Font color | RGB color, black, white |
| Background color | Auto adjust, change manually |

| Data entry | Typing, voice, tapping... |
|---|---|
| Display information | Text, sound |
| Message delivery | Text, voice, alert… |
| Brightness level | Increase/decrease |
| Ring volume | Low, medium, high |
| Sound | Mute, regular, loud |
| Video | On, off |
| .... | .... |

**UI feature set:** A UI feature set is a non-empty set of UI features.
**Disjoint UI feature sets:** Two UI feature sets are said to be disjoint if they have no UI feature in common. It also means that the UI features do not interfere with each other in the process of adaption. For example: {video}, {media sound} and {brightness level} are three disjoint UI feature sets, but {video, media sound} and {video, brightness level} are not disjoint UI feature sets because their intersection is the set {video}.
**Action set:** An action set is a set of actions (also known as values) applied to UI. For example: {Font size is small, Font color is black} is an action set.
**Context category:** We categorized the context information into two categories as shown in Table 3. Physical context information is collected by the mobile device's sensors. The logical information is gathered through the user's registration process, the user's performance, and the user's selections in the setting menu.

Table 3: Context Information Categorization

| Physical Context | local time, local weather (local here also implies the context location considered), device orientation |
|---|---|
| Logical Context | user's profile (age, first time using the app or not, performance) user's preference (color preference, image preference) |

**Context condition:** A context condition is the predicate of the context value. For example: "whether the battery level is low" is a context condition.
**Context set:** A context set is a non-empty set of context. For example: {local time, local weather, device orientation} is a context set.
**Adaption function:**
  Let F be a UI feature set, let C be a context set and let A (C, F) be a function defined over inputs F and C, the output is action set, that is C applied to F.
**Adaption function distributive rule:**
  In A (C, F), let F be divided to disjoint UI feature sets: F = f1 ∪ f2 ∪… ∪ fn, then we have distributive rule:
  A (C, F) = A (C, f1 ∪ f2 ∪…∪ fn,) ≡A (C, f1) ∪A (C, f2) ∪ ....∪ A (C, fn)

Below is an example of different ways of distributing an adaption function:
  ***Problem***: make adaptions on UI features: video, media sound, brightness level, based on context: battery is low.
  ***Solution***:
  1. UI feature set F = {video, media sound, brightness level}
  2. Context set C = {battery is low},
  3. Function: A (C, F) = ({battery is low}, {video, media sound, brightness level})
  4. According to our distributive rule:
     a. If F is divided to disjoint UI feature sets: {video}, {media sound, brightness level}
        Then A ({battery is low}, {video, media sound, brightness level}) = A({battery is low}, {video}) ∪ A({battery is low}, {media sound, brightness level})

b. If F is divided to disjoint UI feature sets: {video, media sound}, {brightness level}
   Then A ({battery is low}, {video, media sound, brightness level}) = A({battery is low}, {video, media sound, brightness level})= A(({battery is low}, {video, media sound}) ∪ A({battery is low}, {brightness level})
c. If F is divided to disjoint UI feature sets: {video}, {media sound, brightness level}
   Then : A ({battery is low}, {video, media sound, brightness level}) = A({battery is low}, { video, media sound, brightness level}) = A(({battery is low}, { video}) ∪ A({battery is low}, {media sound }) ∪ A(({battery is low}, {brightness level })

**Adaption tree:**

We define decision tree over function A (C, F), and we named the decision tree used in our approach as an adaption tree. An adaption tree consists of two types of nodes: condition node and conclusion node. Table 4 shows the nodes and their descriptions in an adaption tree.

Table 4: Node in the Adaption Tree

| Name | Shape | Description |
| --- | --- | --- |
| Condition Node | 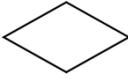 | non-leaf node, denoted as a diamond. It checks the context conditions. |
| Conclusion Node | 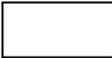 | leaf node, denoted as a rectangle. It represents a UI action after context-based adaption. |

Drawn from top to down, each path from root node to leaf node represents an adaption rule. Each condition node (also recognized as non-leaf node) is represented as a diamond, labeled with the context it checked, and it has several branches coming out of it. Each conclusion node (also recognized as leaf node) is denoted as rectangle, and labeled with UI action.

## 5. The Adaption Tree in the Arithmetic Game Application

We are presenting the adaption rules by constructing an adaption tree in the arithmetic game application. There are various contexts that we can gather. We only collected contexts that could result in at least one action over our application UI.

1. UI feature set: F = {font color, background color, button font color, button background color, weather icon, background image, orientation mode}
   If we divided F to disjoint UI feature sets: F = F1 ∪ F2 = {font color, background color, button font color, button background color, weather icon, background image} ∪ {orientation mode} and name F1 as Theme which is a UI feature set, then we have  F = Theme ∪ {orientation mode}
2. Context set: C = {first time, last unit accuracy, device orientation, local time, local weather} ∪ User's Preference, the User's Preference = {font color preference, background color preference, button font color preference, button background color preference, background image preference}
3. Function: A (C, F) = A ({first time, last unit accuracy, device orientation, local time, local weather, User's Preference}, Theme ∪ {orientation mode})
4. According to our distributive rule:
       A ({first time, last unit accuracy, device orientation, local time, local weather, User's Preference}, Theme ∪ {orientation mode })

≡ A ({first time, last unit accuracy, device orientation, local time, local weather, User's Preference}, Theme)
∪ A ({first time, last unit accuracy, device orientation, local time, local weather, User's Preference }, {orientation mode })

Then define the output:
1. A ({first time, last unit accuracy, device orientation, local time, local weather, User's Preference}, Theme) has three discrete output values: Default Theme, Preferred Color Theme, Weather & time based Theme.
    a. Default Theme = {font color is black, background color is white, button font color is black, button background color is white, weather icon is null, background image is null}
    b. Preferred Color Theme = {font color is A (User's Preference, {font color}), background color is A (User's Preference, {background color}), button font color is A (User's Preference, {button font color}), button background color is A (User's Preference, {button background color}), weather icon is null, background image is null}
    c. Weather & time based Theme = {font color is A (User's Preference, {font color}), background color is A (User's Preference, {background color}), button font color is A (User's Preference, {button font color}), button background color is A (User's Preference, {button background color}), weather icon is A ({local weather} ∪ User's Preference, {weather icon}), background image is A ({local time} ∪ User's Preference, {background image})}
2. A ({first time, last unit accuracy, device orientation, local time, local weather} ∪ User's Preference, {orientation mode}) has two discrete output values: {portrait mode} and {landscape mode}

Figure 2 is an adaption tree over A ({first time, last unit accuracy, device orientation, local time, local weather, User's Preference}, Theme).

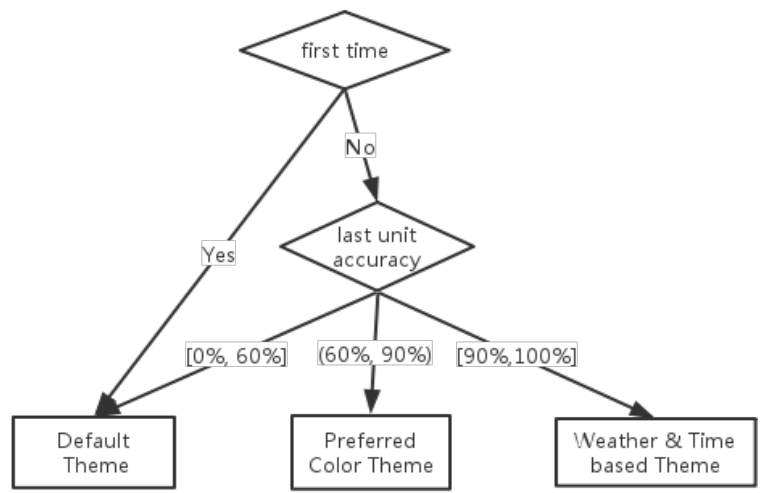

Figure 2: Adaption Tree for Theme

Adaption Tree for Theme

Figure 2 shows four adaption rules for theme:
1. If a user is using this app for the first time, then the theme action is the default color theme.

2. If a user is not using this app for the first time, and the last unit accuracy is between 0% and 60% both inclusive, then the theme action is the default theme.
3. If a user is not using this app for the first time, and the last unit accuracy is between 60% and 90% both exclusive, then the theme action is the preferred color theme.
4. If a user is not using this app for the first time, and the last unit accuracy is between 90% and 100% both inclusive, then the theme action is the weather & time based theme.

If the fourth adaption rule for theme is met (the theme action is the weather & time based theme.), then we can construct an adaption tree for font color, background color, button font color, button background, weather icon, background image to show our rule for those UI features, respectively. Among those UI features, we selected the background image to present our approach.

Figure 4 shows our adaption tree for background image that is based on the adaption tree for theme.

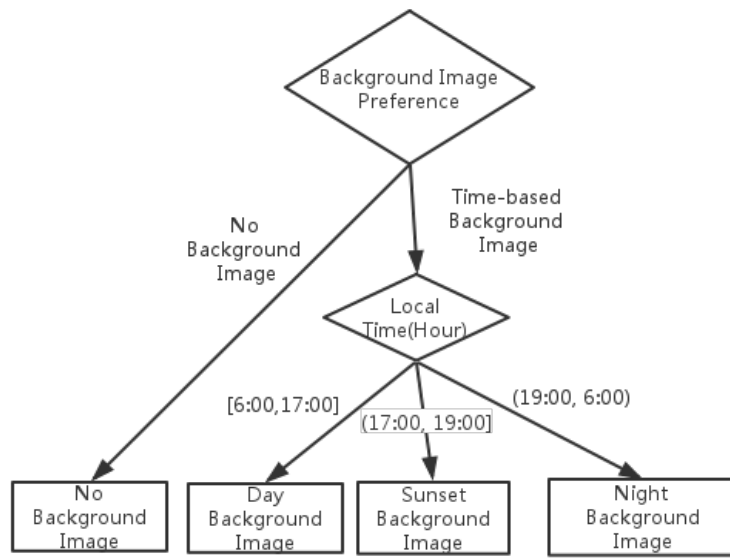

Figure 3: Adaption Tree for Background Image

Figure 3 should be only being considered after Figure 2. The contexts in Figure 2 have higher priority than the contexts in Figure 3.

Figure 2 and Figure 3 together present lots of rules. For example, if the fourth adaption rule for theme is met, the background image preference is a time-based background image, and the local time is between 17:00 to 19:00, then the background image is a sunset background image.

The implementation of the arithmetic game strictly followed the adaption trees in Figures 2 and 3. Below are the screen shots for the application with different themes and with local weather and time information as well.

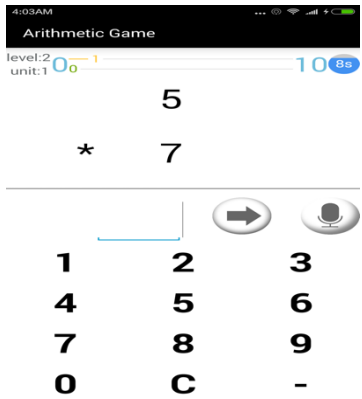 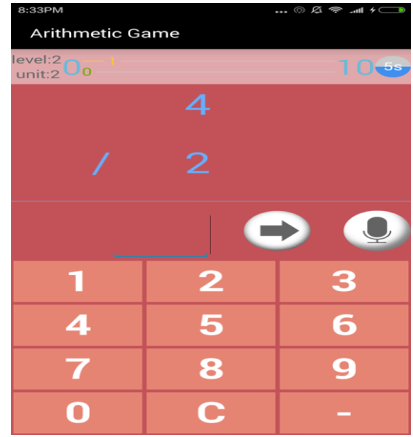

Figure 4 Black and White Theme    Figure 5 User's Preferred Colors

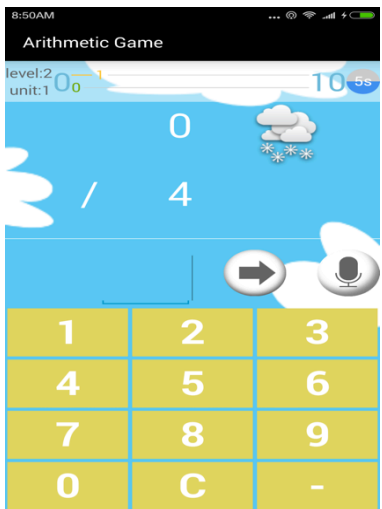 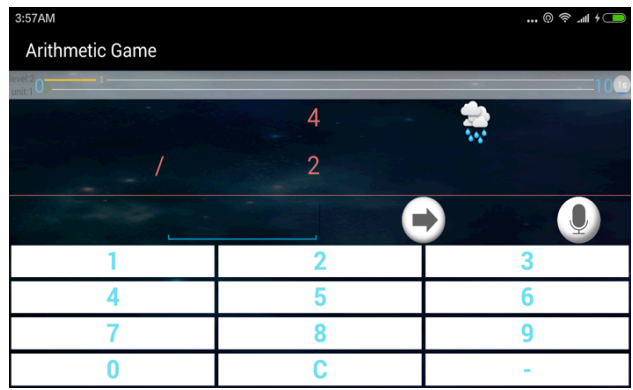

Figure 6  Snowy Day    Figure 7  Rainy Night

## 5. Conclusions

With ubiquitous computing, users access their applications in a wide variety of environments. To cope with various and dynamic execution environments, the adaptive mobile user interface is desired to enhance human-computer interactions. This research work is our attempt to address this issue. We used the rule-based approach, represented as adaption tree to describe the adaption rule for the mobile user interface based on the various context information. Our implementations strictly followed our proposed approach.

It is important to point out we are separating how context is acquired from how it is used, by adapting mobile user interface features to various context information. The user, as a composite entity, is part of the context.

Each context-aware application has its own set of behaviors to react to context modifications. Hence, every software engineer needs to clearly understand the goal of the development and categorize the context in the application. We have proven this idea in two different context-aware applications [16].

The contributions of this research work lie in 1) considering both the user's domain and mobile technology experience in context, 2) detailed modeling inclusion on both input and output data, 3) using the rule to present acquired knowledge in the application. The adaption built into a mobile user interface can enhance the accessibility in the e-commerce domain. The additional benefits are a) increased usability. For example, if the mobile user interface only supports one interaction model, such as typing or voice input/sound output, the usability of the service would be drastically decreased. b) increased awareness of social ethics, e.g. in a quiet room after midnight, the sound could be turned off automatically. c) improved workflow productivity because the mobile user interface is automatically adapted to the dynamic environments.

## Authors

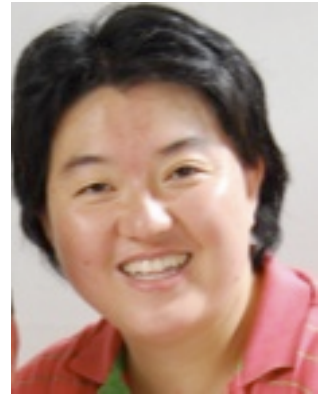

Dr. Mao Zheng is a tenured faculty member and associated professor in the Department of Computer Science at the University of Wisconsin-La Crosse. Her areas of research include Software Engineering, Software Testing and Formal Methods. The courses she has taught include Software Engineering, Software Design, Object-Oriented Development, Software Testing and Object-Oriented Software Development in Java. Dr. Zheng received her Ph.D. in Computer Science at Concordia University in Montreal Canada in 2002. Dr. Zheng has been actively involved with IEEE conferences and served as reviewers and co-organizers for some IEEE conferences. She also has served as an editorial board member for an international journal.

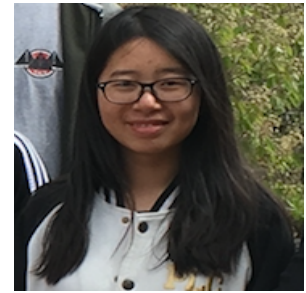

Qian Xu is currently working in Amazon in Seattle USA. She obtained the Mater of Software Engineering degree from the University of Wisconsin – La Crosse in USA in May 2016.

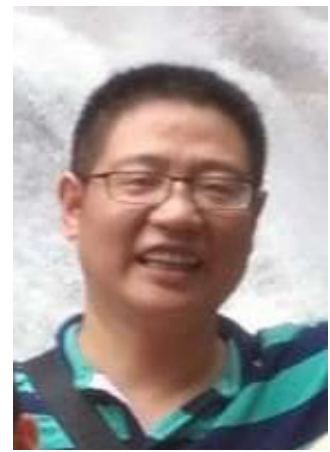

Dr. Hao Fan is a professor at the School of Information Management in Wuhan University in China. His research areas include Data Mining, Data Representation and Management, and Software Engineering.